%% file: 4U1642-49.tex
\documentclass[twocolumn]{aastex631}

\usepackage{gensymb}
\usepackage{amsmath}
\usepackage{xspace}
\usepackage{color}
\usepackage{url}
\usepackage{hyperref}
\usepackage{graphicx}

\shorttitle{X-ray polarimetry of 4U 1624$-$49}
\shortauthors{Saade et al.}

\begin{document}

\title{X-Ray Polarimetry of the Dipping Accreting Neutron Star 4U 1624$-$49}

\author[0000-0001-7163-7015]{M. Lynne Saade}
\affiliation{Science \& Technology Institute, Universities Space Research Association, 320 Sparkman Drive, Huntsville, AL 35805, USA}

\author[0000-0002-3638-0637]{Philip Kaaret}
\affiliation{NASA Marshall Space Flight Center, Huntsville, AL 35812, USA}

\author[0000-0002-0642-1135]{Andrea Gnarini}
\affiliation{Dipartimento di Matematica e Fisica, Universit\`{a} degli Studi Roma Tre, Via della Vasca Navale 84, 00146 Roma, Italy}

\author[0000-0002-0983-0049]{Juri Poutanen}
\affiliation{Department of Physics and Astronomy, 20014 University of Turku, Finland}

\author[0000-0001-9442-7897]{Francesco Ursini}
\affiliation{Dipartimento di Matematica e Fisica, Universit\`{a} degli Studi Roma Tre, Via della Vasca Navale 84, 00146 Roma, Italy}

\author[0000-0002-4622-4240]{Stefano Bianchi}
\affiliation{Dipartimento di Matematica e Fisica, Universit\`{a} degli Studi Roma Tre, Via della Vasca Navale 84, 00146 Roma, Italy}

\author[0009-0009-3183-9742]{Anna Bobrikova}
\affiliation{Department of Physics and Astronomy,  20014 University of Turku, Finland}

\author[0000-0001-8916-4156]{Fabio La Monaca}
\affiliation{INAF Istituto di Astrofisica e Planetologia Spaziali, Via del Fosso del Cavaliere 100, 00133 Roma, Italy}
\affiliation{Dipartimento di Fisica, Universit\`{a} degli Studi di Roma ``Tor Vergata'', Via della Ricerca Scientifica 1, 00133 Roma, Italy}
\affiliation{Dipartimento di Fisica, Universit\`{a} degli Studi di Roma ``La Sapienza'', Piazzale Aldo Moro 5, 00185 Roma, Italy}

\author[0000-0003-0331-3259]{Alessandro Di Marco}
\affiliation{INAF Istituto di Astrofisica e Planetologia Spaziali, Via del Fosso del Cavaliere 100, 00133 Roma, Italy}

\author[0000-0002-6384-3027]{Fiamma Capitanio}
\affiliation{INAF Istituto di Astrofisica e Planetologia Spaziali, Via del Fosso del Cavaliere 100, 00133 Roma, Italy}

\author[0000-0002-5767-7253]{Alexandra Veledina}
\affiliation{Department of Physics and Astronomy,  20014 University of Turku, Finland}
\affiliation{Nordita, KTH Royal Institute of Technology and Stockholm University, Hannes Alfv\'ens v\"ag 12, SE-10691 Stockholm, Sweden}

\include{tier2}

\correspondingauthor{M. Lynne Saade}
\email{mlsaade@usra.edu}

\begin{abstract}
We present the first X-ray polarimetric study of the dipping accreting neutron star 4U~1624$-$49 with the Imaging X-ray Polarimetry Explorer (IXPE). We report a detection of polarization in the non-dip time intervals with a confidence level of 99.99\%. We find an average polarization degree (PD) of $3.1\%\pm0.7\%$ and a polarization angle of $81\degr\pm6\degr$ east of north in the 2--8 keV band. We report an upper limit on the PD of 22\% during the X-ray dips with 95\% confidence. The PD increases with energy, reaching from $3.0\%\pm0.9\%$ in the 4--6 keV band to $6\%\pm2\%$ in the 6--8 keV band. This indicates the polarization likely arises from Comptonization. The high PD observed is unlikely to be produced by Comptonization in the boundary layer or spreading layer alone. It can be produced by the addition of an extended geometrically thin slab corona covering part of the accretion disk, as assumed in previous models of dippers, and/or a reflection component from the accretion disk.
\end{abstract}

\section{Introduction}

Neutron star low-mass X-ray binary (NS--LMXB) spectra contain a hard component believed to be caused by inverse Compton scattering in a population of hot electrons known as the corona. However, the location and size of the corona is a matter of debate. One hypothesis, termed the Eastern model, attributes the corona to a compact region near or on the NS where the accretion disk makes contact with the star \citep[e.g.,][]{1984PASJ...36..741M, 1999AstL...25..269I}. The other primary hypothesis, the Western model, assumes the corona is large and extended \citep[e.g.][]{1988ApJ...324..363W}. These models cannot be distinguished by the means of spectroscopy alone; they both provide adequate fits to medium-resolution NS--LMXB spectra (\citealt{2007ApJ...667.1073L,2011A&A...529A.155C}; and see \citealt{2001AdSpR..28..307B} for a review). X-ray polarimetry can help break the degeneracy between these models by constraining the geometry of the Comptonizing region. 

The Imaging X-ray Polarimetry Explorer (IXPE) enables such measurements. Previous IXPE observations provided constraints on accretion geometries in an initial sample of weakly-magnetized (non-pulsating) NS--LMXBs \mbox{Cyg~X-2} \citep{2023MNRAS.519.3681F}, \mbox{GS 1826$-$238} \citep{2023ApJ...943..129C}, \mbox{XTE J1701$-$462} \citep{2023A&A...674L..10C}, \mbox{GX~9+9} \citep{2023A&A...676A..20U}, \mbox{4U 1820$-$303} \citep{2023ApJ...953L..22D},  \mbox{GX~5$-$1} \citep{2023arXiv231006788F}, 
\mbox{Cir X-1} \citep{Rankin2023}, and \mbox{Sco X-1} \citep{LaMonaca2023}. X-ray polarization has also been measured for \mbox{Sco X-1} \citep{2022ApJ...924L..13L} using Polar Light \citep{2019ExA....47..225F}. In these sources the polarization degree (PD) in the 2--8 keV band has ranged from 1.0\% to 4.6\% whenever detected, and an upper limit of 1.3\% was reported for  GS~1826$-$238. The PD generally increases with increasing energy, and can be interpreted as being attributed to the harder spectral component. Since the Comptonization component is thought to be the harder component of the accreting NS spectrum, these results suggest that the Comptonization component is responsible for most of the polarization. In the sources with a radio jet (\mbox{Sco X-1}, \mbox{Cyg~X-2}, and \mbox{Cir X-1}) the X-ray electric vector position angle (aka polarization angle, PA) 
was sometimes found to be aligned with the jet \citep{2022ApJ...924L..13L,2023MNRAS.519.3681F}, though in some instances it was not \citep{LaMonaca2023,Rankin2023}. The instances where it has been aligned with the jet have been interpreted as evidence of the corona being located in a boundary or spreading layer close to the NS, rather than an extended atmosphere of the disk \citep{2022ApJ...924L..13L}. Thus far, then, X-ray polarization measurements have primarily supported the Eastern model in which the corona is small and similar in size to the NS,  $\sim 10^{6}$ cm.

In contrast, the subset of NS--LMXBs known as dippers has been used to support the Western model where the corona is large and extended \citep[e.g.,][]{2004MNRAS.348..955C}. Dippers show irregular but repeated dips in their X-ray light curves most commonly thought to be caused by obscuration of the X-ray emitting region by clouds occurring in a thickened part of the outer accretion disk, where the accretion stream from the companion star impacts the disk. The spectrum during the dips is thought to arise from a Comptonization component that is only becomes partially obscured while the thermal component becomes completely obscured during the deepest part of dipping. The obscuration of the thermal component is rapid and complete, suggesting it is compact. In contrast, the Comptonization component's obscuration is gradual, with ingress or egress times lasting $\sim$100~s or more. The ingress time $\Delta t$ can be estimated as the time that the leading edge of the absorbing cloud takes to cross the corona due to orbital motion, $\Delta t = P_{\rm orb}(r_{\rm c}/\pi r_{\rm D})$ where $r_{\rm c}$ is the radius of the corona, $r_{\rm D}$ is the outer radius of the disk, and $P_{\rm orb}$ is the orbital period \citep{2004MNRAS.348..955C}.

X-ray dip studies have lead to estimates of Comptonization corona sizes ranging from $\mathrm{3\times10^{9}\;cm}$ to $\mathrm{6\times10^{10}\;cm}$ \citep{2004MNRAS.348..955C}. The corona covers a significant fraction of the accretion disk, from $\sim$6\% to as high as $\sim$50\%, with the size of the corona increasing with increasing luminosity. These size estimates are orders of magnitude larger than those inferred in the Eastern model. Also, Fourier-frequency resolved X-ray spectroscopy indicates that the Comptonization component varies on timescales of milliseconds \citep{2003A&A...410..217G,2013MNRAS.434.2355R}. This can only occur if the hard component is very small, as it would be if it were a boundary layer. The interpretation in terms of an extended corona for the dippers are thus in tension with evidence for small NS coronae.

It should be noted that the aforementioned explanation of dippers is not universally accepted. The reason for the extended corona assumption is that some soft emission persists in the dipping state, which is not consistent with a simple increase in neutral absorption. However, \citet{2005A&A...436..195B} show that the dipping behavior of 4U~1323$-$62 can also be explained by a decrease in the ionization of an absorber, without the need to assume an extended corona. Further exploration of an ionization explanation for the dips is presented for several dipping sources in \citet{2006A&A...445..179D}, including 4U 1624$-$49. The spectral changes of these sources during dipping can be accurately modeled by allowing an increase in column density and a decrease in the ionization state of a heavily ionized absorber. This raises the possibility that dippers have small coronae located in the boundary or spreading layer.

In summary, a variety of NS--LMXBs have been studied with IXPE, and findings have been consistent with the boundary/spreading layer hypothesis for the location of the corona. Here, we investigate the dipper 4U 1624$-$49  to see if these trends continue at higher inclinations.

\section{4U 1624-49}

4U 1624$-$49 is a persistent NS--LMXB. It displays deep \citep[$\sim$ 75\% with respect to 2--10 keV persistent emission;][]{2007A&A...463..289I}, 6--8~h long periodic dips in its light curve that repeat every $\sim$21~h \citep{Watson1985}. Because of these dips, it is expected to have a high inclination \citep[inclination $>$ 60\degr;][]{1987A&A...178..137F}, with the dips created by obscuration of the central source by a thickened part of the outer accretion disk where the incoming stream from the binary companion first enters the accretion disk \citep[e.g.,][]{1982ApJ...253L..61W}. Based on Rossi X-ray Timing Explorer (RXTE) observations, \citet{2005A&A...435.1005L} argued that 4U~1624$-$49 is an atoll source in the banana state, similar to GX~3+1 and GX~9+9, and viewed at a higher inclination. Its Eddington ratio is high for an atoll source \citep[0.5--0.8 $L_{\rm Edd}$;][]{2005A&A...435.1005L}. Unlike other objects of its class, no {
\bf quasi-periodic oscillations} have been observed in 4U 1624$-$49 \citep{2001ApJ...550..962S,2005A&A...435.1005L}. When not in a dip state it sometimes shows flaring behavior of up to 30\% above the persistent 2--10 keV flux level \citep{1989ESASP.296..439J}. The flares are taken as evidence for the LMXB's NS nature \citep{1995A&A...300..441C} and \citet{2001A&A...378..847B} argued they are composed of many rapid individual X-ray bursts.

The earliest attempt to fit the spectrum of 4U 1624$-$49  used a blackbody for the high-flux (non-dip) states and thermal bremsstrahlung for the low-flux (dip) states \citep{Watson1985}. \citet{1989ESASP.296..439J} used a single absorbed power law to fit the non-dip spectrum, while a two-component model consisting of a power-law component and a bremsstrahlung component was used to fit the dip spectrum. Flaring states were well-fit by a blackbody component with $kT\sim2.2$~keV. The dip spectra showed complex absorption that a single absorption component could not fit, with a column density of $\mathrm{\sim2\times10^{24}\,cm^{-2}}$ at the base of the dip. 

\citet{1995A&A...300..441C} proposed the model that became standard in the future literature on 4U 1624$-$49: a two-component model with a blackbody representing the NS, and a power-law component with a cutoff at high energies representing an extended accretion disk corona. In their model, the absorption on the blackbody increases to extremely high levels at the time of deepest dipping, rendering it invisible, while the absorption on the power-law component is nearly constant. A fraction of the power law remains visible, and during saturated dipping, the spectrum is almost entirely a power law.

More analyses of the spectrum were done by \citet{2000A&A...360..583B} and \citet{2001ApJ...550..962S}. They used the basic model from \citet{1995A&A...300..441C} but added additional absorption components. In particular, they added a partial covering absorber with varying column density and covering fraction. This specifically partially covered the extended power-law component when the system was in a dip state, while the blackbody component was completely blocked. In their model, thick obscuring clouds found in the outer edge of the disk progressively cover the extended corona but do not completely hide it from the line of sight, while they completely block the NS from view in saturated dipping. Based on these results, \citet{2000A&A...360..583B} report a radius of $\mathrm{5.3\times10^{10}\;cm}$ for the corona, while 
\citet{2001ApJ...550..962S} report a radius of $\mathrm{5.0\times10^{10}\;cm}$.

A Chandra High-Energy Transmission Grating Spectrometer (HETG) observation of 4U~1624$-$49 was presented in \citet{2007A&A...463..289I}. Both a broad iron line and narrow absorption features were present in the spectrum. The broadening of the line was interpreted as potential evidence of relativistic reflection off of the accretion disk, but they also noted it could be due to Compton scattering in the corona of the accretion disk.

An exploration of 4U~1624$-$49's spectrum using an alternate dipping model was presented by \citet{2006A&A...445..179D}. They investigated the ionization state of the absorption lines in the dipper's spectrum using the XMM-Newton \mbox{EPIC pn} camera. Like \citet{1995A&A...300..441C} they model the non-dip continuum of 4U~1624$-$49 with a blackbody plus a power-law model. However, they include additional neutral and ionized absorption that vary during dipping. The column density of the neutral and ionized absorption increase dramatically as the dipping increases in depth, reaching a maximum of $\mathrm{\sim59\times10^{22}\,cm^{-2}}$ for the neutral absorber and $\mathrm{\sim68\times10^{22}\,cm^{-2}}$ for the ionized one. In contrast to the other dippers studied in the paper, the ionization parameter of the ionized absorber in 4U~1624$-$49 remains constant during dipping. In this picture, thick clouds of neutral and ionized gas in the outer parts of the disk obscure the blackbody component and the corona, but there is no need for the corona to specifically be extended and partially covered.

\begin{figure*}
\centering
\includegraphics[width=\textwidth]{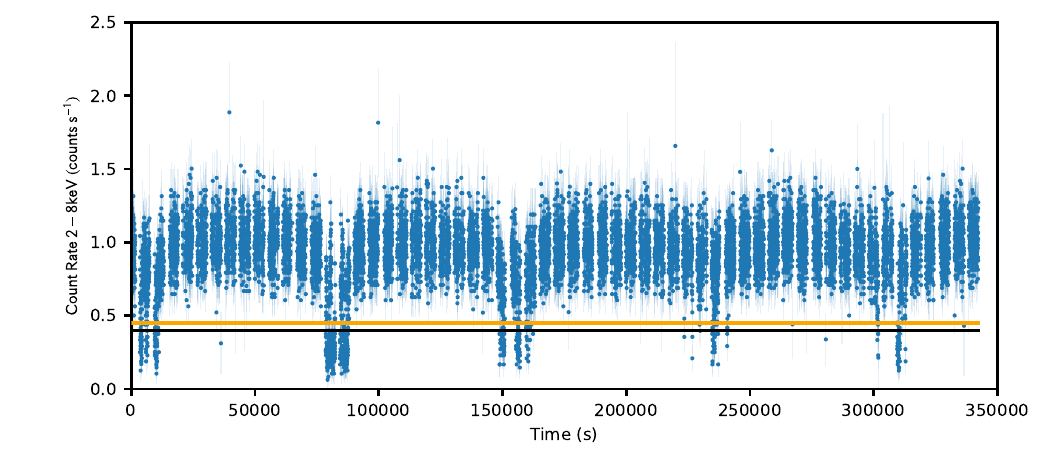}
\caption{IXPE 2--8 keV light curve for the entire observation. The dip times were selected when the count rate was less than 0.4 $\mathrm{counts\;s^{-1}}$. This threshold is marked with a black line. The non-dip times were selected when the count rate was greater than 0.45 $\mathrm{counts\;s^{-1}}$. This threshold is marked with an orange line. The bin size on the light curve is 16 seconds, which is the same used to do the time filtering. The error bars have  been made translucent so the individual points are more easily distinguished.}
\label{fig:light_curve}
\end{figure*}

\begin{figure*}
\centering
\includegraphics[width=0.9\textwidth]{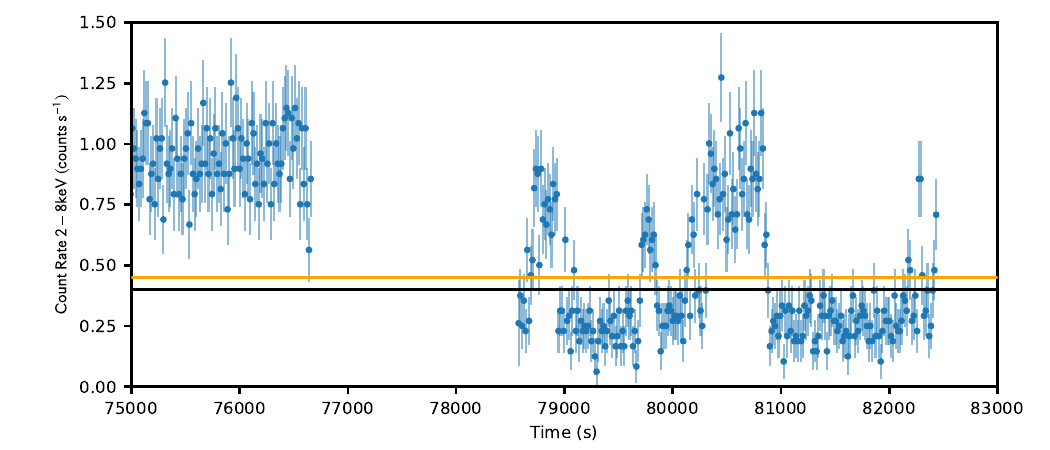}
\caption{Same as Fig.~\ref{fig:light_curve}, but zoomed into the second observed dip and the IXPE orbit immediately preceding it to illustrate the thresholds used for dip and non-dip times selection. 
}
\label{fig:threshold}
\end{figure*}

The most recent spectral study of 4U~1624$-$49 was done by 
\citet{2009ApJ...701..984X}. They found that while a blackbody plus power law fits the spectrum well, a single power law with $\Gamma=2.25$ partially (71\%) covered by local absorption of column density $(\mathrm{8.1^{+0.7}_{-0.6}) \times 10^{22}\; cm^{-2}}$ in addition to the Galactic absorption fits the continuum emission slightly better. Based on variability studies of absorption lines conducted with the Chandra High Energy Transmission Grating Spectrometer they conclude that there is evidence of a two-temperature absorber in the system, a hotter component associated with the corona and a cooler one associated with the accretion disk rim. In their interpretation of the 4U~1624$-$49 system, the corona comprises the inner region of the accretion disk out to a radius $R\sim3\times10^{10}$\,cm and the hot $T\sim3\times10^{6}$\,K gas sandwiches it. The outer accretion disk has a somewhat cooler warm $T\sim1.0\times10^{6}$\;K gas sandwiching it. An absorbing bulge lies at the point on the accretion disk rim where the accretion stream intersects the disk, and a dip occurs when our line of sight passes through it, intersecting the absorbing bulge, the warm gas, and the hot gas around the corona.

With the exception of \citet{2006A&A...445..179D}, these previous analyses of 4U 1624$-$49 conclude that its corona is flat \citep[height-radius ratio of 0.1 at the outer edge;][]{2001ApJ...550..962S} and extended \citep[over 50\% of the accretion disk radius;][]{2001A&A...378..847B}. They have been used as evidence to support the Western model. This contrast with more recent developments which favor the Eastern model motivates X-ray polarization studies of dippers and selection of 4U~1624$-$49 as an IXPE target. While polarization alone cannot determine the size of the corona, the different corona shapes predicted by the Eastern and Western models (the boundary layer of the neutron star vs a slablike geometrically thin atmosphere of the accretion disk) predict different polarization trends with energy, and these trends can be compared to the data obtained by IXPE \citep[e.g.][]{2022MNRAS.514.2561G}.

\section{Observations and Data Analysis}

IXPE \citep{Weisskopf2022} is a NASA/ASI Small Explorer-class mission that was launched on 2021 December 9. It uses three gas-pixel detectors (termed detector units or DU's) to simultaneously collect spatial, temporal, spectral, and polarimetric information. It is most sensitive in the 2--8 keV range and can measure 3 of the 4 Stokes parameters: $I$, $Q$, and $U$.

IXPE observed 4U~1624$-$49 on 2023 August 19 to 23 (ObsID:02007301) for an exposure time of 199~ks. The level-2 event data were analyzed using \textsc{ixpeobssim} v. 30.6.3 \citep{Baldini22}. The source region was a 120\arcsec{} radius circle. The average count rate was 1.17 count\,s$^{-1}$ per DU. We did not subtract the background in the IXPE data, following the prescription reported in \citet{DiMarco2023} for a relatively bright source. 

The entire light curve of the IXPE observation is shown in Figure \ref{fig:light_curve} with 16~s binning. Five dips are clearly visible in the light curve, but no flares are observed. We separated the observation into dip and non-dip times using the total count rate in all three DUs. We chose a threshold of $
\mathrm{<0.40\;counts\;s^{-1}}$ to define the times when the source was in the dipping state, and a threshold of $\mathrm{>0.45\;counts\;s^{-1}}$ to define the times when the source was in a non-dipping state. This was done on the basis of the 16~s bin lightcurve, and these thresholds are plotted over the lightcurve in Figure \ref{fig:light_curve}. A representative illustration of the thresholds used is presented in Figure~\ref{fig:threshold} using the second dip observed by IXPE. We filtered the level-2 event files to only include the source region using \texttt{xpselect}, then filtered them according to the above thresholds using \texttt{xselect}.

\begin{figure}
\centering
\includegraphics[width=\linewidth]{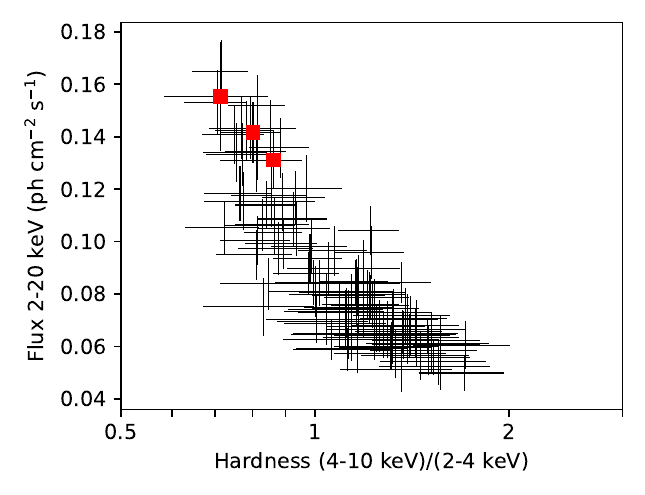}
\caption{Hardness-intensity diagram (HID) from MAXI. Points during the IXPE observation are marked in red.}
\label{fig:hid}
\end{figure}

\begin{deluxetable*}{lccccc}
\tablecaption{Results of Polarimetric Analysis for Different Energy Bands. \label{tab:poltable}}
\tablewidth{0pt}
\tablehead{
\colhead{Quantity} & 
\colhead{2--8 keV} &
\colhead{2--4 keV} &
\colhead{4--6 keV} &
\colhead{6--8 keV} &
\colhead{4--8 keV}}
\startdata
$Q/I$  (\%) & $-2.9\pm0.7$ & $-1.2\pm0.8$ & $-2.8\pm0.9$ & $-6.6\pm2.0$ & $-4.0\pm0.9$\\
$U/I$  (\%) & $0.9\pm0.7$ & $1.0\pm{0.8}$ & $0.6\pm0.8$ & $1.5\pm2.0$ & $0.9\pm0.9$\\
PD  (\%) & $3.1\pm0.7$ & $1.6\pm0.8$ & $2.9\pm0.9$ & $6.4\pm2.0$ & $4.1\pm0.9$\\
PA  (deg) & $81\pm6$ & $71\pm14$ & $84\pm9$ & $83\pm9$ & $84\pm6$
\enddata
\tablecomments{Errors correspond to 1$\sigma$ or {\bf 68\%} CL. 
}
\end{deluxetable*}

\begin{figure}
\centering
\includegraphics[width=\linewidth]{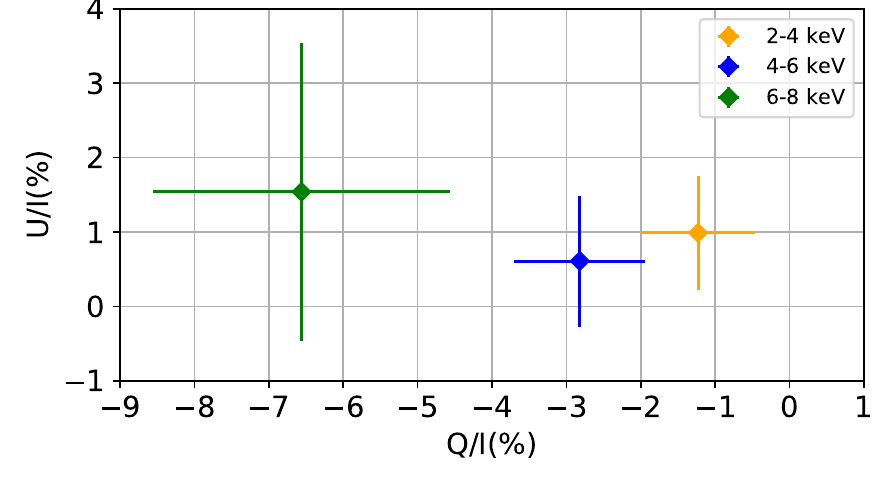}
\caption{Normalized $Q/I$ and $U/I$ Stokes parameters in 2\,keV-wide energy bins as obtained with \texttt{pcube} algorithm.}
\label{fig:stokes_pcube}
\end{figure}

Data from the Monitor of All-Sky X-Ray Image (MAXI)\footnote{http://maxi.riken.jp/} instrument \citep{Matsuoka2009} were used to construct a hardness-intensity diagram (HID) to characterize the source state during the IXPE observations (see Figure~\ref{fig:hid}). The horizontal axis is the hardness ratio, which is defined as the ratio of fluxes in the 4--10~keV versus 2--4~keV bands. The vertical axis is the flux in the 2--20~keV band. Data were taken in the interval MJD 59900 to 60200. Points with a fractional error on the hardness greater than 0.2 were removed for clarity. The flux during the IXPE observations (the red points in Figure \ref{fig:hid}) was higher than average and the X-ray spectrum was softer than average.

\section{Polarimetric Analysis}

\begin{figure}
\centering
\includegraphics[width=\linewidth]{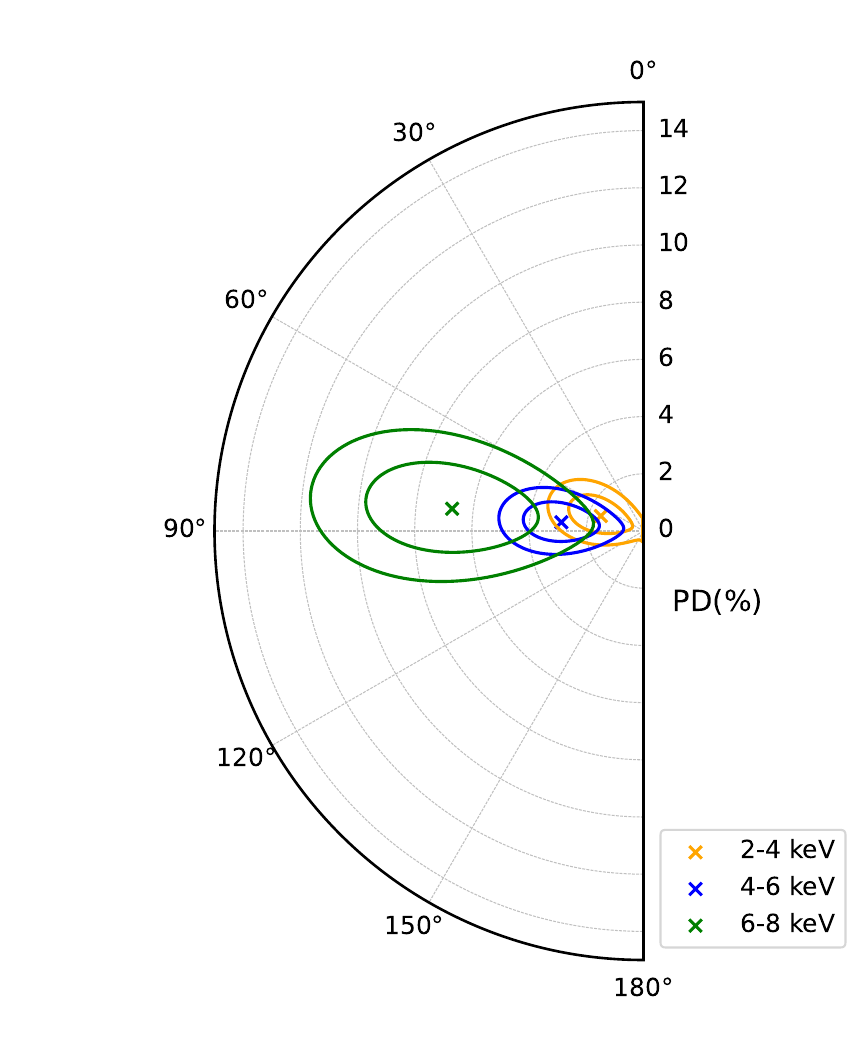}
\caption{Polar plot of PD versus PA contours in 2\,keV-wide energy binning. The contours represent {\bf 68\%} and {\bf 95\%} CL, or $1\sigma$ and $2\sigma$.}
\label{fig:contours_2keV}
\end{figure}

The PD and PA were measured in both the non-dip and dip times using the \texttt{pcube} algorithm \citep{Baldini22}. A highly significant detection of polarization was made in the whole 2--8 keV band for the non-dip times with a confidence level (CL) of 99.99\%. The measured PD was $3.1\% \pm 0.7\%$ and the PA was $81\degr\pm6\degr$ (east of north). Figure~\ref{fig:stokes_pcube} shows the normalized Stokes parameters in the 2\,keV-wide energy bins. The corresponding PD and PA contour plots  are shown in Figure~\ref{fig:contours_2keV}. We also report the values of the normalized Stokes parameters, the PA, and PA in various energy ranges in Table \ref{tab:poltable}. Polarization is significantly detected above 4~keV, while the significance is just 88\% in the 2--4 keV band. We place a 2$\sigma$ upper limit of 3.5\% on the 2--4 keV PD. 

We attempted to measure the polarization during the dip times. The count rate during these times was too low to make a significant detection. We place a 95\% confidence or 2$\sigma$ upper limit of 22\% on the PD during the dip times.

There is a tentative increase of PD with energy that can be seen in Figures \ref{fig:stokes_pcube} and  \ref{fig:contours_2keV} and in Table \ref{tab:poltable}. In contrast the PA does not show a clear trend with energy. We calculated the $\chi^2$/d.o.f. for the assumption that the PD was constant and found a $\chi^2$/d.o.f. of 6.1/2. We find the marginally significant increase of PD with energy at a confidence level of {\bf 95\%}. We can compare the difference in polarization between the 2--4 and 6--8 keV bands, and find a probability of 2.6\% that they would be as far apart as they are by chance.

\begin{figure}
\centering
\includegraphics[width=\columnwidth]{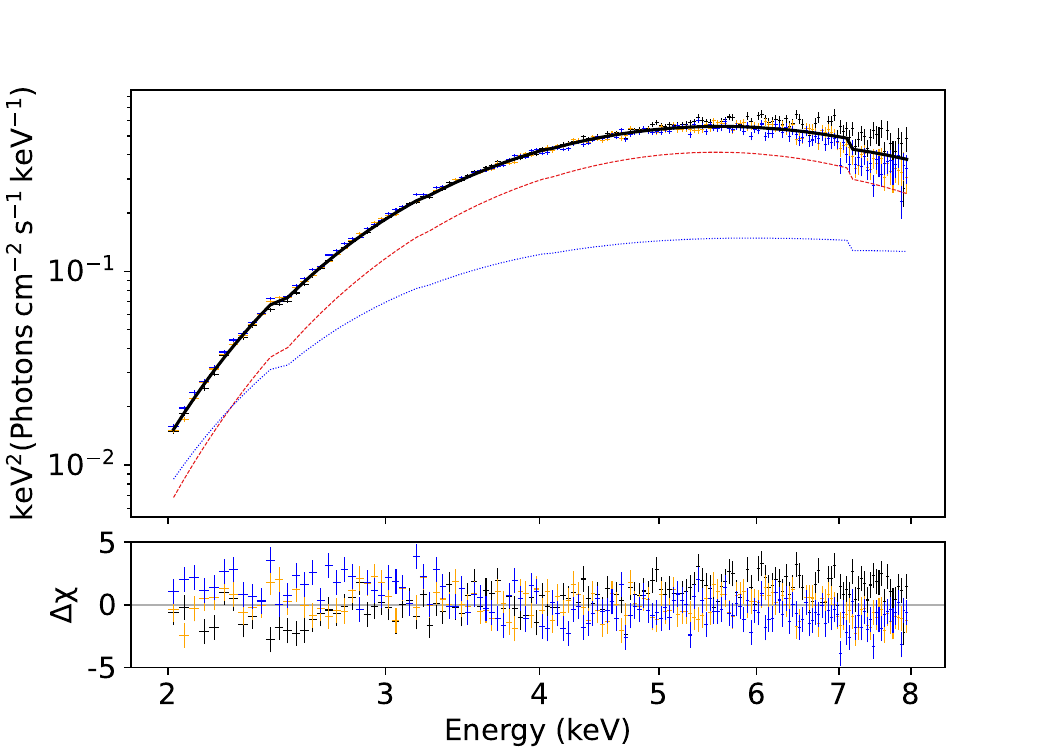}
\caption{Data and unfolded $I$ spectra in $EF_E$ representation for the non-dip intervals using the best-fit \texttt{constant*tbabs*(bbody+cutoffpl)} model. Data from DU1, DU2, and DU3 are shown with black, orange, and blue error bars, respectively. The solid  line is the overall spectral model, the dashed red line is the \texttt{bbody} component, and the dotted blue line is the \texttt{cutoffpl} component. At 5 keV, the stronger component is the \texttt{bbody} component.}
\label{fig:I_spectra}
\end{figure}

\section{Spectropolarimetric Analysis}

\subsection{Western Model}

\begin{figure*}
\centering
\includegraphics[width=\columnwidth]{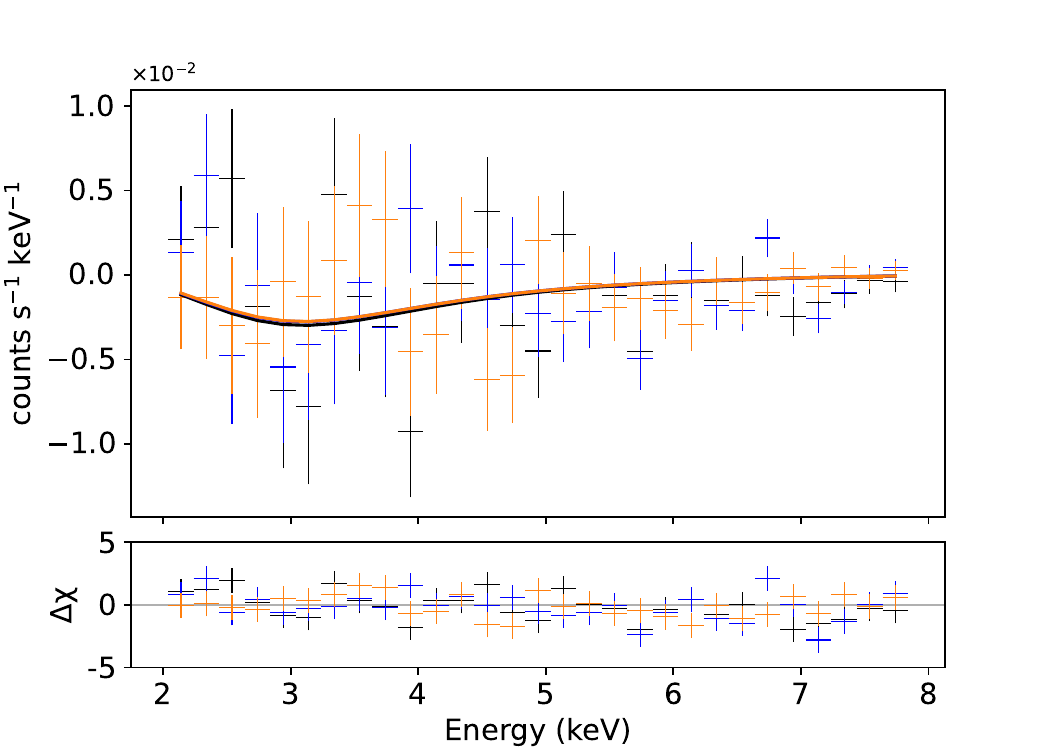}
\includegraphics[width=\columnwidth]{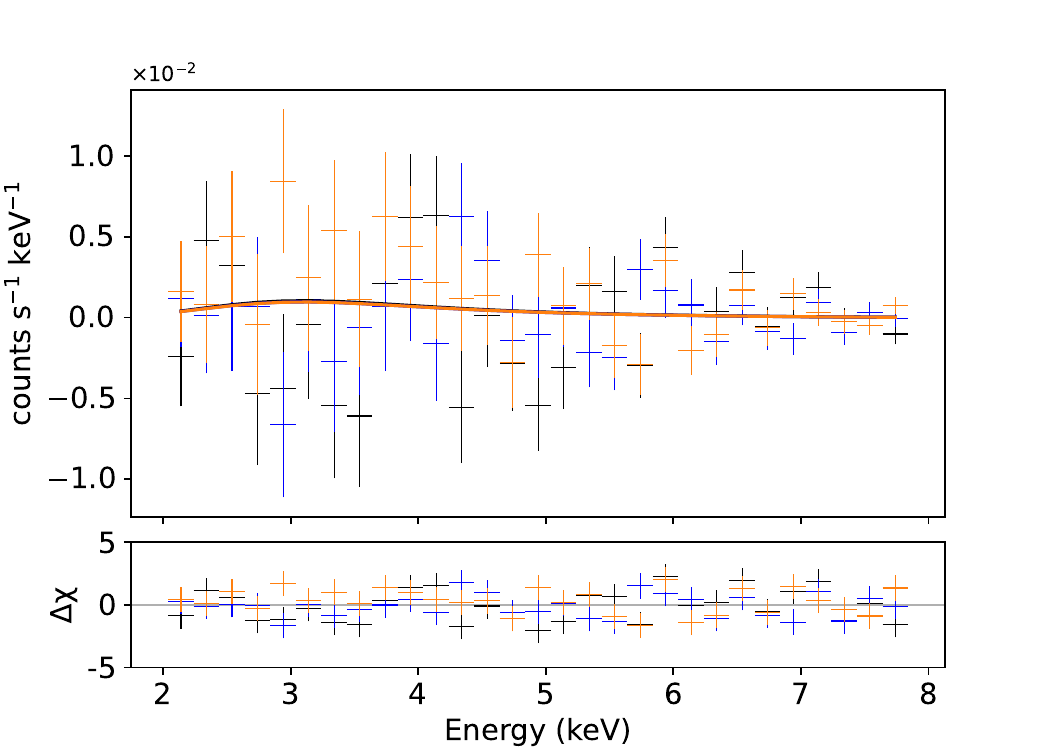}
\caption{Stokes $Q$ (left) and $U$ (right) spectra and the best-fit spectropolarimetric model assuming that only \texttt{cutoffpl} component is  polarized, i.e.  \texttt{constant*tbabs*(bbody+polconst*cutoffpl)}, plotted on a linear scale. The error bars are the data points and the solid line is the overall model. Black, blue, and orange correspond to DU1, DU2, and DU3, respectively.}
\label{fig:QU_spectra}
\end{figure*}

We used \textsc{xspec} version 12.13.1 \citep{1996ASPC..101...17A} for spectropolarimetric analysis. We started by fitting the non-dip $I$ spectra alone with a version of the Western model. We used a \texttt{constant*tbabs*(bbody+cutoffpl)} model following \citet{2000A&A...360..583B}, where \texttt{constant} is the instrumental cross-normalization constant between the three DUs, \texttt{tbabs} represents Galactic absorption, \texttt{bbody} represents blackbody emission from the neutron star, and \texttt{cutoffpl} represents the Comptonized emission from the extended corona. We note that the extension of the \texttt{cutoffpl} to energies below that of the soft photon source would be unphysical. However, in the Western model, the soft photons are thought to arise from the accretion disk which likely has a temperature below the 2~keV lower energy bound of IXPE. We use the model of \citet{2000A&A...360..583B} to enable direct comparision with previously published results. We froze the \texttt{cutoffpl} $\Gamma$ and $E_{\rm cut}$ to the best-fit values from \citet{2000A&A...360..583B} as the limited energy range of IXPE did not allow these parameters to be constrained well. The resulting $\chi^2$/d.o.f. = 749/441. The relatively high $\chi^2$/d.o.f. of 1.7 is likely caused by inconsistencies between the three DUs. The residuals in the fit plotted in Figure \ref{fig:I_spectra} do not show any systematic trends. If we add a systematic error of 3\%, we get a reduced $\chi^2$ of approximately 1 (450/447). The derived spectral parameters should be interpreted with caution. However, the main goal of our spectropolarimetric analysis is the polarization results. The model provides a reasonable fit to the data adequate for extraction of polarization results.

\begin{deluxetable}{lcc}
\tablecaption{Parameters of Best-fit \texttt{constant*tbabs*(bbody+cutoffpl)} Model to Non-dip $I$ Spectrum. \label{tab:I_spectra}}
\tablewidth{0pt}
\tablehead{
\colhead{Component} & 
\colhead{Parameter} &
\colhead{Value}} 
\startdata
\texttt{constant} & $\mathrm{C_{DU1}}$ &  1.0\tablenotemark{a}\\
{} & $\mathrm{C_{DU2}}$ &  $0.96\pm0.01$\\
{} & $\mathrm{C_{DU3}}$ &   $0.92\pm0.01$\\
 \texttt{tbabs} (Galactic) & $N_{\rm H}$ ($\mathrm{10^{22}\;cm^{-2}}$) & $7.7\pm0.3$\\
 \texttt{bbody} & $kT$ (keV) & $1.22\pm{0.01}$\\
 {} & Norm &   $(1.4\pm0.1)\times10^{-2}$\\
  {} & { Flux\tablenotemark{b}} &  { $9.1$}\\
\texttt{cutoffpl} & $\Gamma$ &   2.00\tablenotemark{a}\\
{} & $E_{\rm cut}$ (keV) & 12.0\tablenotemark{a}\\
{} & Norm &   $0.30^{+0.07}_{-0.08}$\\
{} & {Flux\tablenotemark{b}} &  { $4.6$}\\
\hline
{ Total} & { Observed DU1 flux\tablenotemark{b}} & { 7.5}\\
{} & { Unabsorbed flux\tablenotemark{b}} & { 13.7} \\
{} & { Luminosity (erg\;s$^{-1}$)} & { $3.7\times10^{37}$}
\enddata
\tablecomments{Errors correspond to the 90\% CL. }
\tablenotetext{a}{Frozen at this value.}
\tablenotetext{b}{Fluxes are in units of $10^{-10}\;\mathrm{erg\;cm^{-2}\;s^{-1}}$.}
\end{deluxetable}

The parameters of this fit are given in Table~\ref{tab:I_spectra}. The spectrum and model are plotted in Figure~\ref{fig:I_spectra}. The level of Galactic absorption is high, with $N_{\rm H}=(7.7\pm0.3)\times10^{22}$\,cm$^{-2}$, consistent with previously measured $8.6\pm1.0$ by \citet{2000A&A...360..583B}, $8.6\pm3.1$ in  \citet{2001ApJ...550..962S},   and $7.6\pm0.4$ in \citet{2009ApJ...701..984X}, all in units of $10^{22}$\,cm$^{-2}$. The temperature of the \texttt{bbody} component is estimated to be  $1.22\pm{0.01}$ keV which is close to the value of $1.31\pm{0.07}$ from \citet{2000A&A...360..583B} and the blackbody component dominates over the cutoff power-law component across the most of the 2--8~keV band which is consistent with fig.~7 in \citet{2000A&A...360..583B}. Thus, the spectral shape is similar and fixing the powerlaw index and cutoff energy appears reasonable. The \texttt{bbody} temperature is somewhat lower than the $1.52\pm{0.05}$ from \citet{2001ApJ...550..962S} and the $1.39^{+0.05}_{-0.04}$ keV from \citet{2009ApJ...701..984X}. The 2--8 keV luminosity of the source is $3.7\times10^{37}\mathrm{erg\;s^{-1}}$ assuming the distance of 15~kpc \citep{2007ApJ...660.1309X}.

\begin{deluxetable}{lcccc}
\tablecaption{Results of Spectropolarimetric Analysis for Western \texttt{constant*tbabs*(bbody+cutoffpl)} Model.\label{tab:specpol}}
\tablewidth{0pt}
\tablehead{
\colhead{Components} & 
\colhead{} &
\colhead{\texttt{polconst*bbody}} &
\colhead{\texttt{polconst*cutoffpl}}}
\startdata
\texttt{bbody} & PD (\%) & {$3.5\pm{1.3}$} & {0}\\
{} & PA (deg) & {$81\pm11$} & \nodata\\
\hline
\texttt{cutoffpl} & PD (\%) & {0} & {$6.8\pm{2.8}$}\\
{} & PA (deg) & \nodata & {$80\pm12$}\\ 
\hline
 {} & $\chi^2$/d.o.f.  & 956/619 & 960/619\\
\enddata
\tablecomments{Errors correspond to the 90\% CL.}
\end{deluxetable}

The final step of the spectropolarimetric fitting was to fit the $I$, $Q$, and $U$ spectra together. We used the previous best-fit model for the $I$ spectra alone and then tested the effects of assuming that only the \texttt{bbody} component is polarized, or only the \texttt{cutoffpl} component was polarized,  or both. 
The results are presented in Table~\ref{tab:specpol}. We see that the two fits with a single polarized component are of comparable quality. The Stokes $Q$ and $U$ spectra with the best-fit model \texttt{polconst*cutoffpl} are shown in Figure~\ref{fig:QU_spectra}. A similar plot for \texttt{polconst*bbody} model looks identical, so we have not shown it here. When the \texttt{bbody} component was assumed to be polarized, the PD was $3.5\%\pm1.3\%$. For the polarized \texttt{cutoffpl} component, the PD was $6.8\%\pm2.8\%$. The similarity in the $\chi^2$/d.o.f. between the two fits means we cannot distinguish which component is polarized. A fit corresponding to the case where both components are polarized was not statistically better and had unconstrained parameters, so it is not shown here.

\begin{deluxetable}{lcc}
\tablecaption{Parameters of Best-fit \texttt{constant*tbabs*(diskbb+comptt)} Model to Non-dip $I$ Spectrum. \label{tab:I_spectra2}}
\tablewidth{0pt}
\tablehead{
\colhead{Component} & 
\colhead{Parameter} &
\colhead{Value}} 
\startdata
\texttt{constant} & $\mathrm{C_{DU1}}$ &  1.0\tablenotemark{a}\\
{} & $\mathrm{C_{DU2}}$ &  $0.96\pm0.01$\\
{} & $\mathrm{C_{DU3}}$ &   $0.92\pm0.01$\\
 \texttt{tbabs} (Galactic) & $N_{\rm H}$ ($\mathrm{10^{22}\;cm^{-2}}$) & $9.3^{+0.8}_{-0.7}$\\
 \texttt{diskbb} & $kT_{\rm in}$ (keV) & $0.51\pm0.05$\\
 {} & Norm &   $1700^{+2300}_{-1000}$\\
 {} & { Flux\tablenotemark{b}} & { 3.5}\\
\texttt{comptt} & Redshift &   0\tablenotemark{a}\\
{} & $kT_{0}$ (keV) & $1.06\pm{0.03}$\\
{} & $kT_{\rm e}$ (keV) & 3.4\tablenotemark{a}\\
{} & $\tau_{\rm p}$ & 5.1\tablenotemark{a}\\
{} & approx & 1.1\tablenotemark{a}\\
{} & Norm &   ${0.112}^{+0.008}_{-0.006}$\\
{} & { Flux\tablenotemark{b}} & { 13.0}\\
\hline
{ Total} & { Observed DU1 flux\tablenotemark{b}} & { 7.5}\\
{} & {Unabsorbed flux\tablenotemark{b}} & { 16.4}\\
{} & { Luminosity (erg\;s$^{-1}$)} & { $4.4\times10^{37}$}\\
\enddata
\tablecomments{Errors correspond to the 90\% CL. }
\tablenotetext{a}{Frozen at this value.}
\tablenotetext{b}{Fluxes are in units of $10^{-10}\;\mathrm{erg\;cm^{-2}\;s^{-1}}$. }
\end{deluxetable}

\begin{deluxetable}{lcccc}
\tablecaption{Results of Spectropolarimetric Analysis for Eastern \texttt{constant*tbabs*(diskbb+comptt)} Model.\label{tab:specpol2}}
\tablewidth{0pt}
\tablehead{
\colhead{Components} & 
\colhead{} &
\colhead{\texttt{polconst*diskbb}} &
\colhead{\texttt{polconst*comptt}}
}
\startdata
\texttt{diskbb} & PD (\%) & {$5.1\pm{4.5}$} & {0}\\
{} & PA (deg) & {$74\pm31$} & \nodata\\
\hline
\texttt{comptt} & PD (\%) & {0} & {$2.9\pm{1.1}$}\\
{} & PA (deg) & \nodata & {$82\pm11$} \\ 
\hline
 {} & $\chi^2$/d.o.f.  & 1018/619 & 1000/619 \\
\enddata
\tablecomments{Errors correspond to the 90\% CL.}
\end{deluxetable}

\begin{figure}
\centering
\includegraphics[width=\columnwidth]{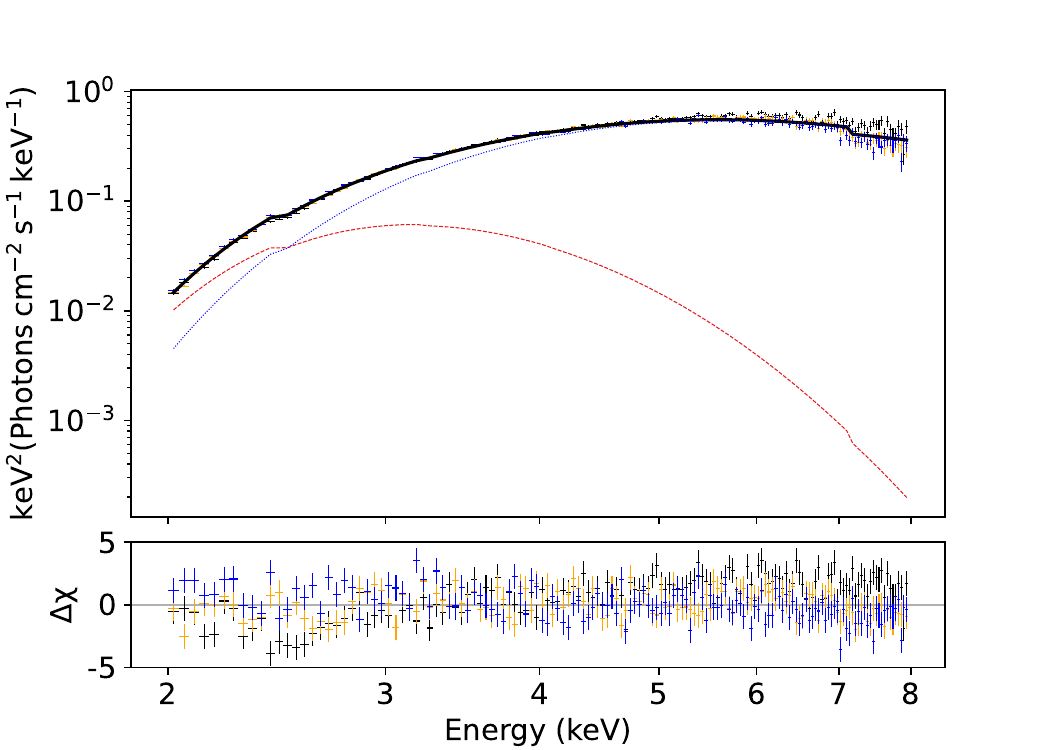}
\caption{Data and unfolded $I$ spectra in $EF_E$ representation for the non-dip intervals using the best-fit \texttt{constant*tbabs*(diskbb+comptt)} model. Data from DU1, DU2, and DU3 are shown with black, orange, and blue error bars, respectively. The solid  line is the overall spectral model, the dashed red line is the \texttt{diskbb} component, and the dotted blue line is the \texttt{comptt} component, which is the stronger one at 5~keV.}
\label{fig:I_spectra2}
\end{figure}

\begin{figure*}
\centering
\includegraphics[width=\columnwidth]{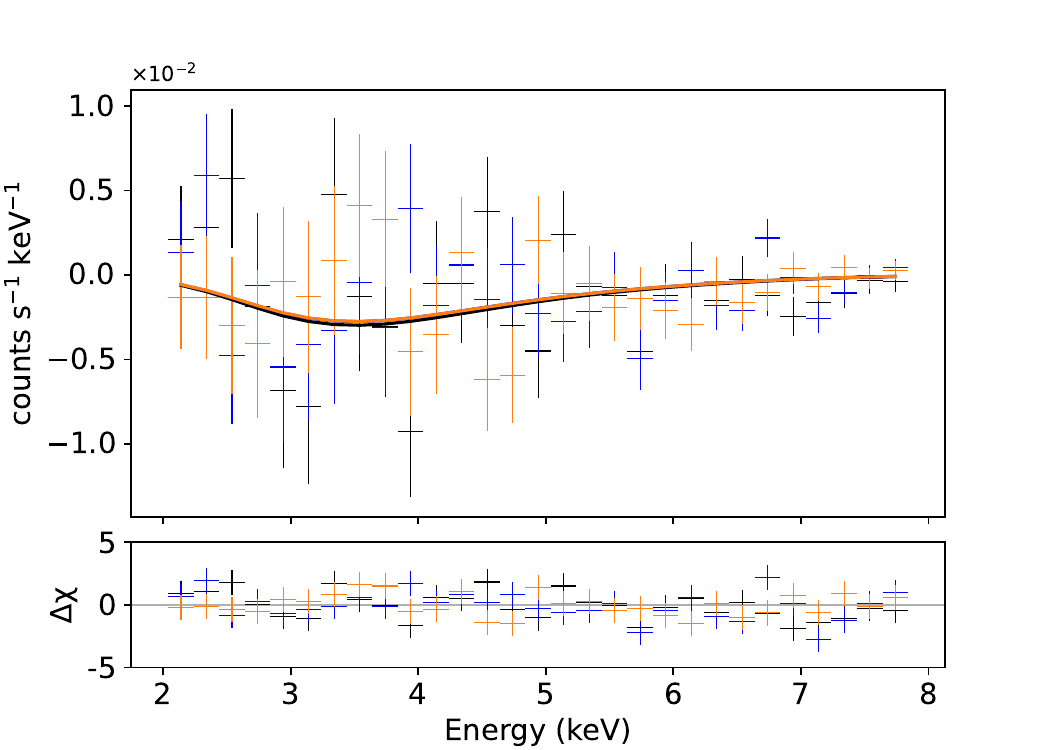}
\includegraphics[width=\columnwidth]{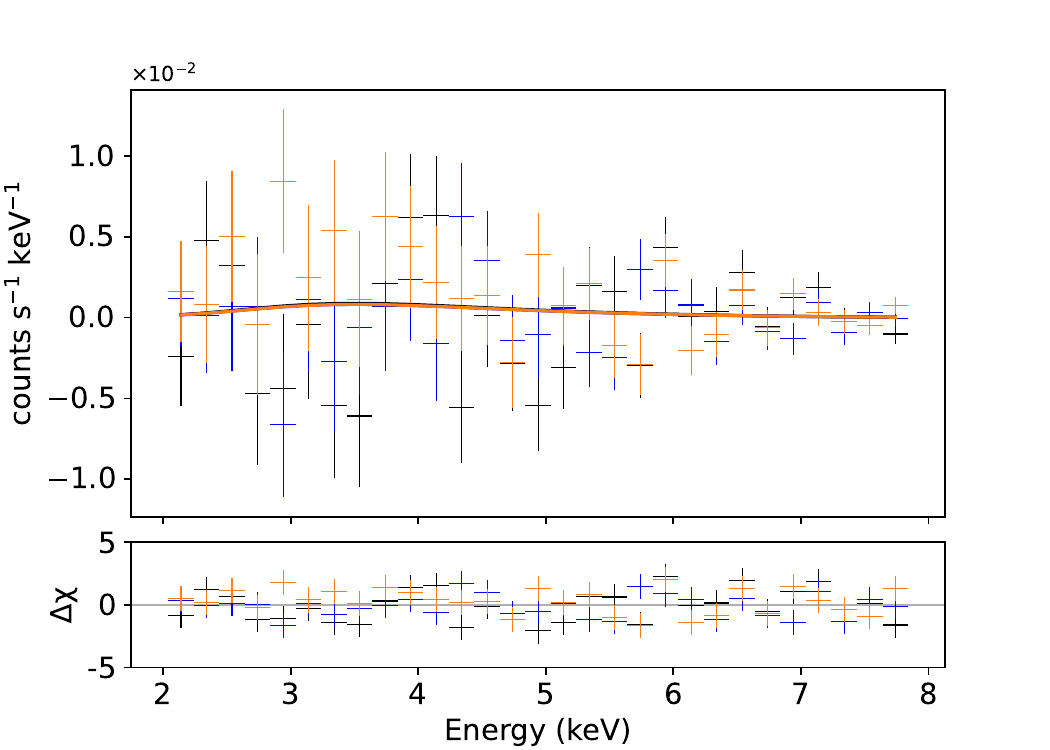}
\caption{Stokes $Q$ (left) and $U$ (right) spectra and the best-fit spectropolarimetric model assuming that only \texttt{comptt} component is  polarized, i.e.  \texttt{constant*tbabs*(diskbb+polconst*comptt)}, plotted on a linear scale. The error bars are the data points and the solid line is the overall model. Black, blue, and orange correspond to DU1, DU2, and DU3, respectively.}
\label{fig:QU_spectra2}
\end{figure*}

\subsection{Eastern Model}

For a representative Eastern model, we chose a \texttt{constant*tbabs*(diskbb+comptt)} model to fit the $I$ spectrum. In this model the \texttt{diskbb} component represents the multitemperature blackbody emission from the accretion disk, while the \texttt{comptt} component represents the Comptonized emission from the boundary or spreading layer. Initial attempts at fitting the entire model at once left most parameters unconstrained, so we started the fitting again with \texttt{constant*tbabs*comptt} model alone. We froze the approx parameter to 1.1 so we were using the spherical geometry (as would be more appropriate for a neutron star boundary layer/spreading layer). We froze the coronal electron temperature $kT_{\rm e}$ and the corona optical depth $\tau_{\rm p}$ to their values from GX~9+9, which were 3.4 and 5.1, respectively \citep{2023A&A...676A..20U}. As noted above, GX~9+9 is thought to have similar properties to 4U~1624$-$49. The best value of the seed photon temperature from this fit was $kT_{0}$=1.1 keV. We then added a \texttt{diskbb} component to the model and set the inner accretion disk temperature $kT_{\rm in}$ to be $<1.1$ keV and the seed photon temperature to be $kT_{0}>1.1$ keV. This lead to $kT_0$ being pegged at 1.1 keV, so we reset $kT_{0}$ to be less than 0.9 keV. The resulting fit had  $\chi^{2}$/d.o.f.=738/440, which is a modest improvement over the Western model fit. Like the Western model fit, if we add a systematic error of 3\%, we get a reduced $\chi^{2}$ close to 1 (440/447).

The parameters of the Eastern model fit are tabulated in Table~\ref{tab:I_spectra2}. The level of Galactic absorption is higher in this fit, $N_{\rm H}=(9.3^{+0.8}_{-0.7})\times10^{22}$\,cm$^{-2}$. The values of $kT_{\rm in}$ and $kT_{0}$ are similar to that of other NS--LMXBs observed with IXPE \citep[e.g.,][]{2023MNRAS.519...50U,2023MNRAS.519.3681F,2023ApJ...943..129C,2023ApJ...953L..22D,Rankin2023}. There is a wide uncertainty in the normalization of the \texttt{diskbb} Component, likely due to the fact most of its emission is below the IXPE band.  The Eastern model fit is plotted over the $I$ spectrum in Figure \ref{fig:I_spectra2}.

We froze the parameters of the $I$ spectrum model before adding polarization components to the model. For the spectropolarimetric fits with this model, we tested two possibilities with either the \texttt{diskbb} component or the \texttt{comptt} component is polarized. The parameters of these fits are shown in Table \ref{tab:specpol2}. We can see the fit with the polarized \texttt{comptt}  component is much better than the fit with the polarized  \texttt{diskbb}. This indicates that in this model, the polarization derives primarily from the Comptonization component. The spectropolarimetric fit for the Eastern model with the polarized \texttt{comptt} component is plotted over the $Q$ and $U$ spectra in Figure~\ref{fig:QU_spectra2}.

We also attempted a fit with both the \texttt{diskbb} and \texttt{comptt} components being polarized, but the PA on the \texttt{diskbb} component was unconstrained in this fit. Even if we froze the \texttt{diskbb} PA to be $90\degr$ offset from the \texttt{comptt} PA (as is expected based on theory, e.g., \citealt{1960ratr.book.....C,1963trt..book.....S}), the \texttt{diskbb} PD was unconstrained. These fits are therefore not explored any further.

\section{Discussion}

The PD of 3.1\% measured for 4U 1624$-$49 in the 2--8 keV band is consistent with the previous X-ray polarization measurements for NS--LMXBs, which range from about 1.0\%  in Sco X-1 \citep{LaMonaca2023} to 4.6\% in XTE J1701$-$462 \citep{2023A&A...674L..10C}. The PD does show marginal significance for an increase with energy as also observed for other NS--LMXBs. A notable example is 4U 1820$-$303, which shows a strong increase in the polarization with energy  \citep{2023ApJ...953L..22D}. The PD strongly depends on both the shape of the Comptonizing region and the inclination of the source with respect to the line of sight \citep[e.g.,][]{2022MNRAS.514.2561G}. The PA of 4U 1624$-$49 cannot be compared with theoretical expectations since the jet direction of the source is unknown, but there is no evidence of its rotation with energy. 

The Western model spectropolarimetric results are ambiguous with regards to whether the polarization is predominantly derived from the soft (i.e. blackbody) component or the hard Comptonization (i.e. cutoff power-law) component. In contrast, in the Eastern model, spectropolarimetric results clearly indicate the polarization is primarily produced by the Comptonization component. The overall reduced $\chi^{2}$ of the Western model fit is somewhat better than for the Eastern model fit.

4U 1624$-$49 has specifically been described as similar to GX~9+9 \citep{2005A&A...435.1005L} in being a bright atoll source that remains in the banana state. The IXPE observation of GX~9+9 suggests that the polarized radiation arises from multiple components, being a combination of Comptonization in a quasi-spherical or wedge-shaped spreading layer plus reflection of soft photons above the accretion disk \citep{2023A&A...676A..20U}.

The increasing PD with energy of 4U 1624$-$49 is consistent with the classical results for a centrally illuminated accretion disk \citep{1993MNRAS.260..663M}. The soft NS photons scatter off the disk and get reflected toward the observer. These reflected photons are highly polarized \citep[up to $\sim$20\%;][]{1993MNRAS.260..663M,1996MNRAS.283..892P} and may significantly contribute to the polarization signal even if their fraction of the total flux is small \citep{2016A&A...596A..21I}. When the spreading layer illuminates the accretion disk, the PD can reach up to 6\% depending on the inclination \citep{1985MNRAS.217..291L}. However, this is overestimated as the direct disk contribution is not taken into account, and it tends to lower the polarization since its PA is perpendicular to that of the reflected radiation \citep{1993MNRAS.260..663M,1996MNRAS.283..892P}. 4U~1624$-$49 shows a broad iron line \citep{2007A&A...463..289I} that is likely due to reflection of X-rays off of the accretion disk. There therefore is most likely a contribution to the polarization from a reflection component.

The increasing PD with energy is also consistent with results of Monte Carlo numerical simulations considering a thick slab/torus covering the inner accretion disk or a wedge-like Comptonizing region close to the NS \citep{2022MNRAS.514.2561G,2023ApJ...943..129C,2023A&A...676A..20U}. However, the observed PD is relatively large compared to the numerical predictions for these geometries. A large PD could be explained if the Comptonizing region is a geometrically thin and optically thick slab covering the accretion disk, in which case the classical results by \cite{1960ratr.book.....C} can be applied. If we assume that the polarized signal is completely due to the atmosphere of the accretion disk, the measured PD of 3.1\% would thus be consistent with a slab at an inclination of 66\degr. It should be noted that the inclination of dippers without eclipses is typically around $\sim65\degr$ \citep{2007A&A...463..289I}. 
 
It is worth comparing our results for 4U~1624$-$49 to the IXPE results for the dipping, black hole candidate X-ray binary 4U~1630$-$47. Like 4U~1624$-$49, 4U~1630$-$47 shows dips in its light curve thought to be caused by obscuration of the inner accretion disk \citep{1998ApJ...494..747T,1998ApJ...494..753K} that indicate it has an inclination $\gtrsim60\degr$ \citep{1998ApJ...494..753K,2013Natur.504..260D}. It was also observed by IXPE in a high soft state \citep{Ratheesh2023} and shows a PD increasing with energy, and a constant PA. However, the PD is higher, $\sim8\%$ in the 2--8 keV band. Unlike 4U~1624$-$49, it does not require a Comptonization component to fit its IXPE and NICER spectra, suggesting its polarization arises entirely from the accretion disk emission or its scattering in the surrounding cold  material, e.g. equatorial wind. In NS--LXMBs like 4U~1624$-$49, contributions from the spreading layer or neutron star surface, perhaps unpolarized or  with polarization orthogonal to the accretion disk, which could lower the observed polarization.

We note that a cooler atmosphere of plasma has been observed from 4U~1624$-49$ \citep{2009ApJ...701..984X}. Even if this plasma does not contribute significantly to the observed Comptonized component of the spectrum, scattering within it still could impart additional polarization to the outgoing light.

Subsequent deeper IXPE observations of dippers, including 4U~1624$-$49, would allow the PD variation with energy to be better constrained, allowing models of corona geometries in these sources to be more accurately tested. It would also allow the upper limits on polarization in the dip states to be more tightly constrained, which could be used to further test the Western model. If the Western model is correct, one could expect the polarization in the dip state to be higher than in the non-dip state, since the presumably less polarized blackbody component would be blocked, in a manner similar to the high polarization observed in the obscured active galaxy Circinus  \citep{2023MNRAS.519...50U}.

If scattering of the central source radiation in a large cool corona (or a wind) is responsible for the observed polarization, then in reality both Eastern and Western models would be correct in some sense: the Eastern model accurately describes the properties of the emission region where most of the energy is produced, while the Western model describes the property of the extended cool corona (or possibly a wind) which scatters radiation from the central source. This is consistent with developments in the field since the original Eastern and Western models were proposed, in which it has been found that the X-ray spectra of LMXBs encode information about multiple independent emission regions, such as the accretion disk and the boundary layer \citep{2007ApJ...667.1073L,2011A&A...529A.155C}.

\section*{Acknowledgments}

IXPE is a joint US and Italian mission.  The US contribution is supported by the National Aeronautics and Space Administration (NASA) and led and managed by its Marshall Space Flight Center (MSFC), with industry partner Ball Aerospace (contract NNM15AA18C).  The Italian contribution is supported by the Italian Space Agency (Agenzia Spaziale Italiana, ASI) through contract ASI-OHBI-2022-13-I.0, agreements ASI-INAF-2022-19-HH.0 and ASI-INFN-2017.13-H0, and its Space Science Data Center (SSDC) with agreements ASI-INAF-2022-14-HH.0 and ASI-INFN 2021-43-HH.0, and by the Istituto Nazionale di Astrofisica (INAF) and the Istituto Nazionale di Fisica Nucleare (INFN) in Italy.  This research used data products provided by the IXPE Team (MSFC, SSDC, INAF, and INFN) and distributed with additional software tools by the High-Energy Astrophysics Science Archive Research Center (HEASARC), at NASA Goddard Space Flight Center (GSFC).
This research has made use of the MAXI data provided by RIKEN, JAXA and the MAXI team.

We acknowledge support from the Academy of Finland grants 333112, 349144, and 355672 (JP, AB, AV, SST) and the German Academic Exchange Service (DAAD) travel grant 57525212 (VD).

\bibliography{4U1642-49}{}
\bibliographystyle{aasjournal}

\end{document}

%% file: tier2.tex
\author[0000-0002-3777-6182]{Iv\'an Agudo}
\affiliation{Instituto de Astrof\'{i}sica de Andaluc\'{i}a -- CSIC, Glorieta de la Astronom\'{i}a s/n, 18008 Granada, Spain}
\author[0000-0002-5037-9034]{Lucio A. Antonelli}
\affiliation{INAF Osservatorio Astronomico di Roma, Via Frascati 33, 00040 Monte Porzio Catone (RM), Italy}
\affiliation{Space Science Data Center, Agenzia Spaziale Italiana, Via del Politecnico snc, 00133 Roma, Italy}
\author[0000-0002-4576-9337]{Matteo Bachetti}
\affiliation{INAF Osservatorio Astronomico di Cagliari, Via della Scienza 5, 09047 Selargius (CA), Italy}
\author[0000-0002-9785-7726]{Luca Baldini}
\affiliation{Istituto Nazionale di Fisica Nucleare, Sezione di Pisa, Largo B. Pontecorvo 3, 56127 Pisa, Italy}
\affiliation{Dipartimento di Fisica, Universit\`{a} di Pisa, Largo B. Pontecorvo 3, 56127 Pisa, Italy}
\author[0000-0002-5106-0463]{Wayne H. Baumgartner}
\affiliation{NASA Marshall Space Flight Center, Huntsville, AL 35812, USA}
\author[0000-0002-2469-7063]{Ronaldo Bellazzini}
\affiliation{Istituto Nazionale di Fisica Nucleare, Sezione di Pisa, Largo B. Pontecorvo 3, 56127 Pisa, Italy}
\author[0000-0002-0901-2097]{Stephen D. Bongiorno}
\affiliation{NASA Marshall Space Flight Center, Huntsville, AL 35812, USA}
\author[0000-0002-4264-1215]{Raffaella Bonino}
\affiliation{Istituto Nazionale di Fisica Nucleare, Sezione di Torino, Via Pietro Giuria 1, 10125 Torino, Italy}
\affiliation{Dipartimento di Fisica, Universit\`{a} degli Studi di Torino, Via Pietro Giuria 1, 10125 Torino, Italy}
\author[0000-0002-9460-1821]{Alessandro Brez}
\affiliation{Istituto Nazionale di Fisica Nucleare, Sezione di Pisa, Largo B. Pontecorvo 3, 56127 Pisa, Italy}
\author[0000-0002-8848-1392]{Niccol\`{o} Bucciantini}
\affiliation{INAF Osservatorio Astrofisico di Arcetri, Largo Enrico Fermi 5, 50125 Firenze, Italy}
\affiliation{Dipartimento di Fisica e Astronomia, Universit\`{a} degli Studi di Firenze, Via Sansone 1, 50019 Sesto Fiorentino (FI), Italy}
\affiliation{Istituto Nazionale di Fisica Nucleare, Sezione di Firenze, Via Sansone 1, 50019 Sesto Fiorentino (FI), Italy}
\author[0000-0003-1111-4292]{Simone Castellano}
\affiliation{Istituto Nazionale di Fisica Nucleare, Sezione di Pisa, Largo B. Pontecorvo 3, 56127 Pisa, Italy}
\author[0000-0001-7150-9638]{Elisabetta Cavazzuti}
\affiliation{Agenzia Spaziale Italiana, Via del Politecnico snc, 00133 Roma, Italy}
\author[0000-0002-4945-5079]{Chien-Ting Chen}
\affiliation{Science \& Technology Institute, Universities Space Research Association, 320 Sparkman Drive, Huntsville, AL 35805, USA}
\author[0000-0002-0712-2479]{Stefano Ciprini}
\affiliation{Istituto Nazionale di Fisica Nucleare, Sezione di Roma ``Tor Vergata'', Via della Ricerca Scientifica 1, 00133 Roma, Italy}
\affiliation{Space Science Data Center, Agenzia Spaziale Italiana, Via del Politecnico snc, 00133 Roma, Italy}
\author[0000-0003-4925-8523]{Enrico Costa}
\affiliation{INAF Istituto di Astrofisica e Planetologia Spaziali, Via del Fosso del Cavaliere 100, 00133 Roma, Italy}
\author[0000-0001-5668-6863]{Alessandra De Rosa}
\affiliation{INAF Istituto di Astrofisica e Planetologia Spaziali, Via del Fosso del Cavaliere 100, 00133 Roma, Italy}
\author[0000-0002-3013-6334]{Ettore Del Monte}
\affiliation{INAF Istituto di Astrofisica e Planetologia Spaziali, Via del Fosso del Cavaliere 100, 00133 Roma, Italy}
\author[0000-0002-5614-5028]{Laura Di Gesu}
\affiliation{Agenzia Spaziale Italiana, Via del Politecnico snc, 00133 Roma, Italy}
\author[0000-0002-7574-1298]{Niccol\`{o} Di Lalla}
\affiliation{Department of Physics and Kavli Institute for Particle Astrophysics and Cosmology, Stanford University, Stanford, California 94305, USA}
\author[0000-0002-4700-4549]{Immacolata Donnarumma}
\affiliation{Agenzia Spaziale Italiana, Via del Politecnico snc, 00133 Roma, Italy}
\author[0000-0001-8162-1105]{Victor Doroshenko}
\affiliation{Institut f\"{u}r Astronomie und Astrophysik, Universit\"{a}t T\"{u}bingen, Sand 1, 72076 T\"{u}bingen, Germany}
\author[0000-0003-0079-1239]{Michal Dov\v{c}iak}
\affiliation{Astronomical Institute of the Czech Academy of Sciences, Bo\v{c}n\'{i} II 1401/1, 14100 Praha 4, Czech Republic}
\author[0000-0003-4420-2838]{Steven R. Ehlert}
\affiliation{NASA Marshall Space Flight Center, Huntsville, AL 35812, USA}
\author[0000-0003-1244-3100]{Teruaki Enoto}
\affiliation{RIKEN Cluster for Pioneering Research, 2-1 Hirosawa, Wako, Saitama 351-0198, Japan}
\author[0000-0001-6096-6710]{Yuri Evangelista}
\affiliation{INAF Istituto di Astrofisica e Planetologia Spaziali, Via del Fosso del Cavaliere 100, 00133 Roma, Italy}
\author[0000-0003-1533-0283]{Sergio Fabiani}
\affiliation{INAF Istituto di Astrofisica e Planetologia Spaziali, Via del Fosso del Cavaliere 100, 00133 Roma, Italy}
\author[0000-0003-1074-8605]{Riccardo Ferrazzoli}
\affiliation{INAF Istituto di Astrofisica e Planetologia Spaziali, Via del Fosso del Cavaliere 100, 00133 Roma, Italy}
\author[0000-0003-3828-2448]{Javier A. Garc\'{i}a}
\affiliation{X-ray Astrophysics Laboratory, NASA Goddard Space Flight Center, Greenbelt, MD 20771, USA}
\author[0000-0002-5881-2445]{Shuichi Gunji}
\affiliation{Yamagata University,1-4-12 Kojirakawa-machi, Yamagata-shi 990-8560, Japan}
\author{Kiyoshi Hayashida}
\altaffiliation{Deceased}
\affiliation{Osaka University, 1-1 Yamadaoka, Suita, Osaka 565-0871, Japan}
\author[0000-0001-9739-367X]{Jeremy Heyl}
\affiliation{University of British Columbia, Vancouver, BC V6T 1Z4, Canada}
\author[0000-0002-0207-9010]{Wataru Iwakiri}
\affiliation{International Center for Hadron Astrophysics, Chiba University, Chiba 263-8522, Japan}
\author[0000-0001-9522-5453]{Svetlana G. Jorstad}
\affiliation{Institute for Astrophysical Research, Boston University, 725 Commonwealth Avenue, Boston, MA 02215, USA}
\affiliation{Department of Astrophysics, St. Petersburg State University, Universitetsky pr. 28, Petrodvoretz, 198504 St. Petersburg, Russia}
\author[0000-0002-5760-0459]{Vladimir Karas}
\affiliation{Astronomical Institute of the Czech Academy of Sciences, Bo\v{c}n\'{i} II 1401/1, 14100 Praha 4, Czech Republic}
\author[0000-0001-7477-0380]{Fabian Kislat}
\affiliation{Department of Physics and Astronomy and Space Science Center, University of New Hampshire, Durham, NH 03824, USA}
\author{Takao Kitaguchi}
\affiliation{RIKEN Cluster for Pioneering Research, 2-1 Hirosawa, Wako, Saitama 351-0198, Japan}
\author[0000-0002-0110-6136]{Jeffery J. Kolodziejczak}
\affiliation{NASA Marshall Space Flight Center, Huntsville, AL 35812, USA}
\author[0000-0002-1084-6507]{Henric Krawczynski}
\affiliation{Physics Department and McDonnell Center for the Space Sciences, Washington University in St. Louis, St. Louis, MO 63130, USA}
\author[0000-0002-0984-1856]{Luca Latronico}
\affiliation{Istituto Nazionale di Fisica Nucleare, Sezione di Torino, Via Pietro Giuria 1, 10125 Torino, Italy}
\author[0000-0001-9200-4006]{Ioannis Liodakis}
\affiliation{NASA Marshall Space Flight Center, Huntsville, AL 35812, USA}
\author[0000-0002-0698-4421]{Simone Maldera}
\affiliation{Istituto Nazionale di Fisica Nucleare, Sezione di Torino, Via Pietro Giuria 1, 10125 Torino, Italy}
\author[0000-0002-0998-4953]{Alberto Manfreda}  
\affiliation{Istituto Nazionale di Fisica Nucleare, Sezione di Napoli, Strada Comunale Cinthia, 80126 Napoli, Italy}
\author[0000-0003-4952-0835]{Fr\'{e}d\'{e}ric Marin}
\affiliation{Universit\'{e} de Strasbourg, CNRS, Observatoire Astronomique de Strasbourg, UMR 7550, 67000 Strasbourg, France}
\author[0000-0002-2055-4946]{Andrea Marinucci}
\affiliation{Agenzia Spaziale Italiana, Via del Politecnico snc, 00133 Roma, Italy}
\author[0000-0001-7396-3332]{Alan P. Marscher}
\affiliation{Institute for Astrophysical Research, Boston University, 725 Commonwealth Avenue, Boston, MA 02215, USA}
\author[0000-0002-6492-1293]{Herman L. Marshall}
\affiliation{MIT Kavli Institute for Astrophysics and Space Research, Massachusetts Institute of Technology, 77 Massachusetts Avenue, Cambridge, MA 02139, USA}
\author[0000-0002-1704-9850]{Francesco Massaro}
\affiliation{Istituto Nazionale di Fisica Nucleare, Sezione di Torino, Via Pietro Giuria 1, 10125 Torino, Italy}
\affiliation{Dipartimento di Fisica, Universit\`{a} degli Studi di Torino, Via Pietro Giuria 1, 10125 Torino, Italy}
\author[0000-0002-2152-0916]{Giorgio Matt}
\affiliation{Dipartimento di Matematica e Fisica, Universit\`{a} degli Studi Roma Tre, Via della Vasca Navale 84, 00146 Roma, Italy}
\author{Ikuyuki Mitsuishi}
\affiliation{Graduate School of Science, Division of Particle and Astrophysical Science, Nagoya University, Furo-cho, Chikusa-ku, Nagoya, Aichi 464-8602, Japan}
\author[0000-0001-7263-0296]{Tsunefumi Mizuno}
\affiliation{Hiroshima Astrophysical Science Center, Hiroshima University, 1-3-1 Kagamiyama, Higashi-Hiroshima, Hiroshima 739-8526, Japan}
\author[0000-0003-3331-3794]{Fabio Muleri}
\affiliation{INAF Istituto di Astrofisica e Planetologia Spaziali, Via del Fosso del Cavaliere 100, 00133 Roma, Italy}	
\author[0000-0002-6548-5622]{Michela Negro} 
\affiliation{Department of Physics and Astronomy, Louisiana State University, Baton Rouge, LA 70803, USA}
\author[0000-0002-5847-2612]{Chi-Yung Ng}
\affiliation{Department of Physics, The University of Hong Kong, Pokfulam, Hong Kong}
\author[0000-0002-1868-8056]{Stephen L. O'Dell}
\affiliation{NASA Marshall Space Flight Center, Huntsville, AL 35812, USA}
\author[0000-0002-5448-7577]{Nicola Omodei}
\affiliation{Department of Physics and Kavli Institute for Particle Astrophysics and Cosmology, Stanford University, Stanford, California 94305, USA}
\author[0000-0001-6194-4601]{Chiara Oppedisano}
\affiliation{Istituto Nazionale di Fisica Nucleare, Sezione di Torino, Via Pietro Giuria 1, 10125 Torino, Italy}
\author[0000-0001-6289-7413]{Alessandro Papitto}
\affiliation{INAF Osservatorio Astronomico di Roma, Via Frascati 33, 00040 Monte Porzio Catone (RM), Italy}
\author[0000-0002-7481-5259]{George G. Pavlov}
\affiliation{Department of Astronomy and Astrophysics, Pennsylvania State University, University Park, PA 16801, USA}
\author[0000-0001-6292-1911]{Abel L. Peirson}
\affiliation{Department of Physics and Kavli Institute for Particle Astrophysics and Cosmology, Stanford University, Stanford, California 94305, USA}
\author[0000-0003-3613-4409]{Matteo Perri}
\affiliation{Space Science Data Center, Agenzia Spaziale Italiana, Via del Politecnico snc, 00133 Roma, Italy}
\affiliation{INAF Osservatorio Astronomico di Roma, Via Frascati 33, 00040 Monte Porzio Catone (RM), Italy}
\author[0000-0003-1790-8018]{Melissa Pesce-Rollins}
\affiliation{Istituto Nazionale di Fisica Nucleare, Sezione di Pisa, Largo B. Pontecorvo 3, 56127 Pisa, Italy}
\author[0000-0001-6061-3480]{Pierre-Olivier Petrucci}
\affiliation{Universit\'{e} Grenoble Alpes, CNRS, IPAG, 38000 Grenoble, France}
\author[0000-0001-7397-8091]{Maura Pilia}
\affiliation{INAF Osservatorio Astronomico di Cagliari, Via della Scienza 5, 09047 Selargius (CA), Italy}
\author[0000-0001-5902-3731]{Andrea Possenti}
\affiliation{INAF Osservatorio Astronomico di Cagliari, Via della Scienza 5, 09047 Selargius (CA), Italy}
\author[0000-0002-2734-7835]{Simonetta Puccetti}
\affiliation{Space Science Data Center, Agenzia Spaziale Italiana, Via del Politecnico snc, 00133 Roma, Italy}
\author[0000-0003-1548-1524]{Brian D. Ramsey}
\affiliation{NASA Marshall Space Flight Center, Huntsville, AL 35812, USA}
\author[0000-0002-9774-0560]{John Rankin}
\affiliation{INAF Istituto di Astrofisica e Planetologia Spaziali, Via del Fosso del Cavaliere 100, 00133 Roma, Italy}
\author[0000-0003-0411-4243]{Ajay Ratheesh}
\affiliation{INAF Istituto di Astrofisica e Planetologia Spaziali, Via del Fosso del Cavaliere 100, 00133 Roma, Italy}
\author[0000-0002-7150-9061]{Oliver J. Roberts}
\affiliation{Science \& Technology Institute, Universities Space Research Association, 320 Sparkman Drive, Huntsville, AL 35805, USA}
\author[0000-0001-6711-3286]{Roger W. Romani}
\affiliation{Department of Physics and Kavli Institute for Particle Astrophysics and Cosmology, Stanford University, Stanford, California 94305, USA}
\author[0000-0001-5676-6214]{Carmelo Sgr\`{o}}
\affiliation{Istituto Nazionale di Fisica Nucleare, Sezione di Pisa, Largo B. Pontecorvo 3, 56127 Pisa, Italy}
\author[0000-0002-6986-6756]{Patrick Slane}
\affiliation{Center for Astrophysics, Harvard \& Smithsonian, 60 Garden St, Cambridge, MA 02138, USA}
\author[0000-0002-7781-4104]{Paolo Soffitta}
\affiliation{INAF Istituto di Astrofisica e Planetologia Spaziali, Via del Fosso del Cavaliere 100, 00133 Roma, Italy}
\author[0000-0003-0802-3453]{Gloria Spandre}
\affiliation{Istituto Nazionale di Fisica Nucleare, Sezione di Pisa, Largo B. Pontecorvo 3, 56127 Pisa, Italy}
\author[0000-0002-2954-4461]{Douglas A. Swartz}
\affiliation{Science \& Technology Institute, Universities Space Research Association, 320 Sparkman Drive, Huntsville, AL 35805, USA}
\author[0000-0002-8801-6263]{Toru Tamagawa}
\affiliation{RIKEN Cluster for Pioneering Research, 2-1 Hirosawa, Wako, Saitama 351-0198, Japan}
\author[0000-0003-0256-0995]{Fabrizio Tavecchio}
\affiliation{INAF Osservatorio Astronomico di Brera, via E. Bianchi 46, 23807 Merate (LC), Italy}
\author[0000-0002-1768-618X]{Roberto Taverna}
\affiliation{Dipartimento di Fisica e Astronomia, Universit\`{a} degli Studi di Padova, Via Marzolo 8, 35131 Padova, Italy}
\author{Yuzuru Tawara}
\affiliation{Graduate School of Science, Division of Particle and Astrophysical Science, Nagoya University, Furo-cho, Chikusa-ku, Nagoya, Aichi 464-8602, Japan}
\author[0000-0002-9443-6774]{Allyn F. Tennant}
\affiliation{NASA Marshall Space Flight Center, Huntsville, AL 35812, USA}
\author[0000-0003-0411-4606]{Nicholas E. Thomas}
\affiliation{NASA Marshall Space Flight Center, Huntsville, AL 35812, USA}
\author[0000-0002-6562-8654]{Francesco Tombesi}
\affiliation{Dipartimento di Fisica, Universit\`{a} degli Studi di Roma ``Tor Vergata'', Via della Ricerca Scientifica 1, 00133 Roma, Italy}
\affiliation{Istituto Nazionale di Fisica Nucleare, Sezione di Roma ``Tor Vergata'', Via della Ricerca Scientifica 1, 00133 Roma, Italy}
\affiliation{Department of Astronomy, University of Maryland, College Park, Maryland 20742, USA}
\author[0000-0002-3180-6002]{Alessio Trois}
\affiliation{INAF Osservatorio Astronomico di Cagliari, Via della Scienza 5, 09047 Selargius (CA), Italy}
\author[0000-0002-9679-0793]{Sergey S. Tsygankov}
\affiliation{Department of Physics and Astronomy,  20014 University of Turku, Finland}
\author[0000-0003-3977-8760]{Roberto Turolla}
\affiliation{Dipartimento di Fisica e Astronomia, Universit\`{a} degli Studi di Padova, Via Marzolo 8, 35131 Padova, Italy}
\affiliation{Mullard Space Science Laboratory, University College London, Holmbury St Mary, Dorking, Surrey RH5 6NT, UK}
\author[0000-0002-4708-4219]{Jacco Vink}
\affiliation{Anton Pannekoek Institute for Astronomy \& GRAPPA, University of Amsterdam, Science Park 904, 1098 XH Amsterdam, The Netherlands}
\author[0000-0002-5270-4240]{Martin C. Weisskopf}
\affiliation{NASA Marshall Space Flight Center, Huntsville, AL 35812, USA}
\author[0000-0002-7568-8765]{Kinwah Wu}
\affiliation{Mullard Space Science Laboratory, University College London, Holmbury St Mary, Dorking, Surrey RH5 6NT, UK}
\author[0000-0002-0105-5826]{Fei Xie}
\affiliation{Guangxi Key Laboratory for Relativistic Astrophysics, School of Physical Science and Technology, Guangxi University, Nanning 530004, China}
\affiliation{INAF Istituto di Astrofisica e Planetologia Spaziali, Via del Fosso del Cavaliere 100, 00133 Roma, Italy}
\author[0000-0001-5326-880X]{Silvia Zane}
\affiliation{Mullard Space Science Laboratory, University College London, Holmbury St Mary, Dorking, Surrey RH5 6NT, UK}